\documentstyle[aps,floats,psfig]{revtex}

\begin{document}
\psfigurepath{.:plot:figure}
\twocolumn[\hsize\textwidth\columnwidth\hsize\csname @twocolumnfalse\endcsname
\bibliographystyle{unsrt}
\preprint{draft:htm6.tex, \today}
\title{Phase transitions in a magnetic field in
V$_{2-y}$O$_3$ ($y=0$ and $0.04$)}
\author{Wei Bao, A. H. Lacerda and J. D. Thompson} 
\address{Los Alamos National Laboratory, Los Alamos, NM 87545}
\author{J. M. Honig and P. Metcalf}
\address{Department of Chemistry, Purdue University, West Lafayette, IN 47907
}
\date{\today}
\maketitle
\begin{abstract}
The effect of a magnetic field on the two long-range-ordered magnetic phases,
the collinear insulating antiferromagnetic (AFI) and the incommensurate
metallic transverse spin density wave (SDW), 
is investigated for the vanadium sesquioxide
system. A field of 18~T has little effect on the AFI phase of a nominal
V$_2$O$_3$ sample. The transverse SDW phase of V$_{1.96}$O$_3$ can 
be suppressed by 
a 4.6(3)-T magnetic field applied in the plane of spiraling spins,
while the same magnetic field applied along the spiral
axis has little effect on the SDW phase.
\end{abstract}
\vskip2pc]

\narrowtext

Vanadium sesquioxide (V$_2$O$_3$) has been extensively investigated, 
as a prototype Mott-Hubbard system exhibiting a 
metal-insulator transition\cite{rev_mit,rev_mnm}.
Stimulated by interest in the Mott-Hubbard system after the
discovery of superconductivity in cuprates, a revisit of this classic
material has generated exciting new findings:
discovery of the static incommensurate spin density wave order\cite{bao93},
spin excitations with completely different spatial correlations 
between the AFI phase and all other phases\cite{bao94a,bao96c,bao96a,bao98a}
orbital occupation with $S=1$ character\cite{park}, 
prominence of orbital contributions to magnetic fluctuations
in high temperature metallic phase\cite{vtaki},
anomalous resonant x-ray scattering in the AFI phase\cite{orb_pao},
and high resolution photoemission\cite{hrpes}. 
These results are inconsistent with previously accepted theoretical
models. One outcome
is a renewed interest in theoretical investigation
on the orbital degree of freedom in correlated $d$-electron
systems\cite{bibrice,orb_yqli,orb_milaf,biboles,orb_eaks}.

\begin{figure}[bt]
\centerline{
\psfig{file=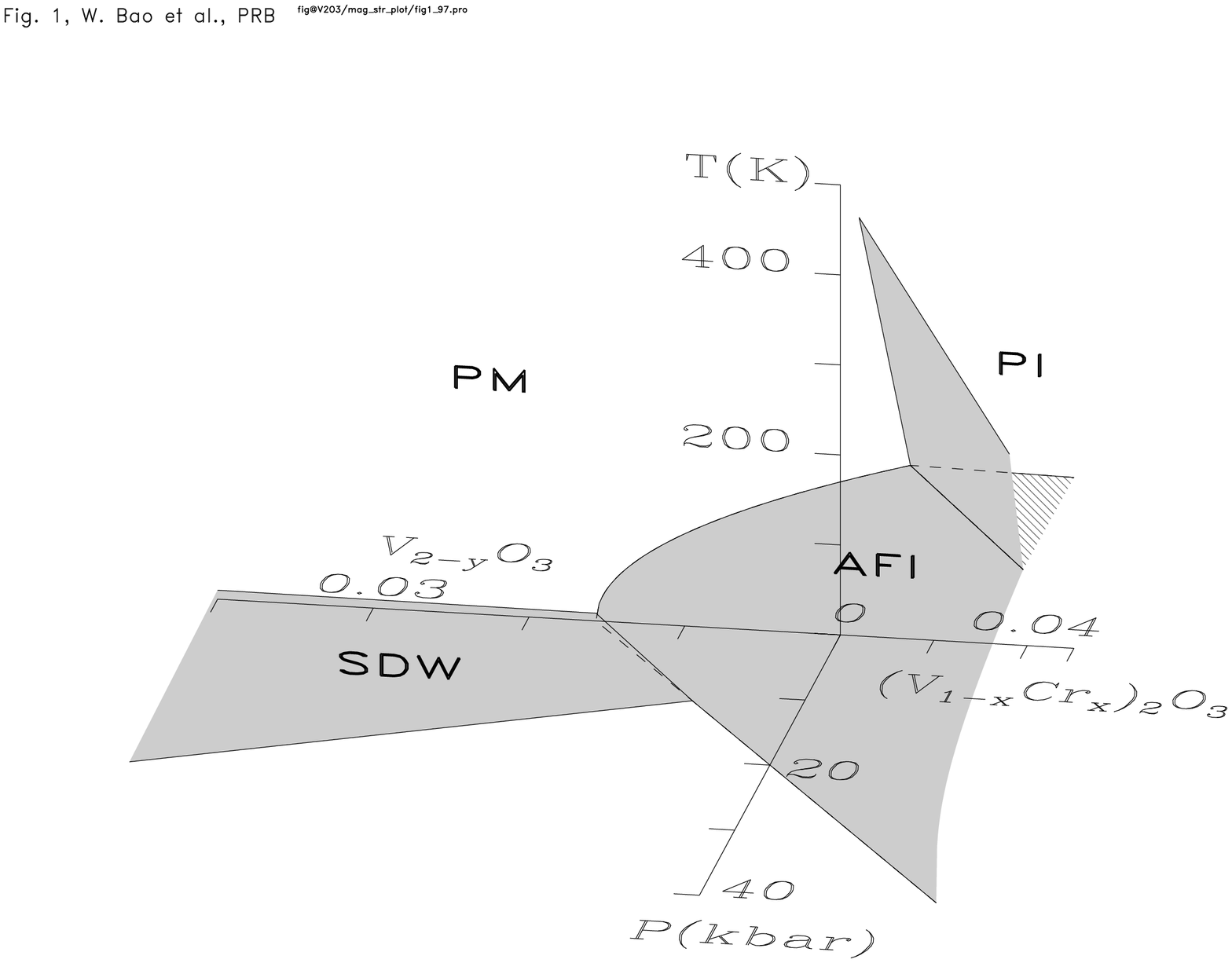,width=\columnwidth,angle=90,clip=}}
\caption{The composition-pressure-temperature phase diagram
for V$_2$O$_3$ system. From Fig.~1 of \protect\cite{bao98a},
based on \protect\cite{bibdbml,bibyukkd,bao93}. Both paramagnetc (P) and antiferromagnetic (AF)
phases occur in insulating (I) and metallic (M) states. The antiferromagnetic
metallic state (AFM) is now known to be a SDW caused by Fermi surface
nesting.}
\label{phs}
\end{figure}

The V$_2$O$_3$ system is very sensitive to external 
perturbation. Studies using high pressure and doping with various
ions have proven fruitful. For example, the anomaly in physical properties 
around 500~K in the pure V$_2$O$_3$ is found to be caused by a nearby
critical point between the paramagnetic metallic (PM) phase and
the paramagnetic insulating (PI) phase, which is uncovered by Cr 
doping\cite{bibdbmb} (refer to Fig.~\ref{phs}).
The AFM phase, discovered with excess oxygen or Ti 
doping\cite{bibyukk}, encourages theoretical 
investigations of this possibility in the Hubbard 
model\cite{mthbcyrl,inftymitb}. 
High pressure studies
reveal that doping has effects beyond only exerting internal chemical
pressure\cite{bibsctfr}. It also allows investigation into low temperature
properties of the paramagnetic metallic phase\cite{bibdbma}, 
which is interrupted
by phase transitions at ambient pressure.

In this work, we explore new phase space of
V$_2$O$_3$ using high magnetic fields.  
The influence of a static magnetic field (up to 18~T)
on the two low temperature magnetic ground states,
the AFI and the SDW, was investigated at ambient pressure.
While a magnetic field of 18~T is too weak to have a
significant
effect on the AFI transition of a V$_2$O$_3$ sample,
it completely suppresses the transverse SDW phase of a V$_{1.96}$O$_3$ sample.


Single crystals of V$_2$O$_3$ were grown using a skull melter\cite{bibsmpl}.
The as-grown crystal used to study the AFI transition has 
a slight oxygen excess, 
which reduced the AFI transition
temperature to T$_C \approx 160$~K.
Another sliced crystal about 1~mm thick was annealed in a suitably chosen 
CO-CO$_2$ atmosphere\cite{bibsmplb} for 2 weeks at 1400$^o$C to
adjust the stoichiometry to V$_{1.96}$O$_3$. This sample has an SDW ground
state at ambient pressure with T$_N = 8.7$~K.

Magnetization was measured using a vibrating sample magnetometer (VSM) in a
20-T superconducting magnet at the Pulsed Field Facility of the National
High Magnetic Field Laboratory at Los Alamos National Laboratory (NHMFL-LANL).
The crystal sample was first wrapped in Teflon tape, and then
attached to the sample holder at the tip of cold finger by Teflon tape. 
In this arrangement, the sample could rotate inside the sample holder
in response to the applied vertical magnetic field.
Reproducible measurements were
obtained after first ramping the field to 18~T, the maximum magnetic
field used in this work. 

\begin{figure}[bt]
\centerline{
\psfig{file=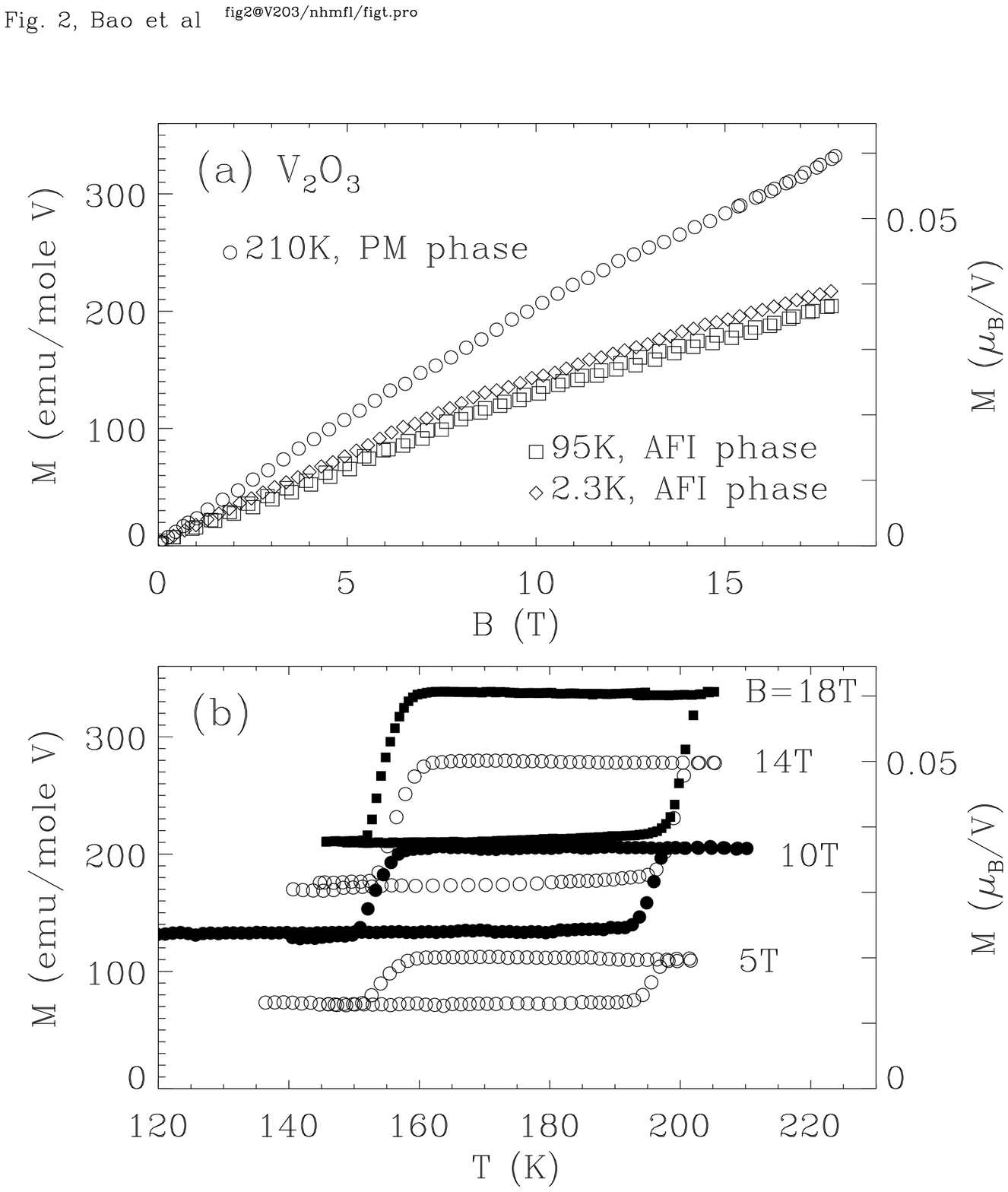,width=\columnwidth,angle=0,clip=}}
\caption{(a) Magnetization curve for V$_2$O$_3$ at 2.3, 95 and 210~K.
(b) Magnetic hysteresis loops in various magnetic fields. 
The AFI transition temperature does not change appreciably in this field range.}
\label{fig_MvH}
\end{figure}

The effect of magnetic field on the AFI phase transition was first
investigated with measurements of 
magnetization as a function of magnetic field
for an as-grown V$_2$O$_3$ crystal at various
temperatures below and above T$_C\approx 160$~K. 
A few examples are shown in Fig.~\ref{fig_MvH}(a).
Data at 95~K were taken with the field both ramping up and down.
The small difference between the field-up and field-down measurements
may not be significant.
Clearly, no first-order phase transition between the PM and AFI
is detectable in the
magnetization curve up to the highest field, $B=18$~T.
In this paper, the $B$ field is always referred to magnetic field
outside the sample.

The AFI phase transition was then investigated with
magnetization as a function of temperature at various field strengths
[refer to Fig.~\ref{fig_MvH}(b)].
Thermal hysteresis, due to the first-order
PM-AFI phase transition, in data taken on warming and cooling
hardly changes with $B$. Thus, we conclude that a magnetic field
of 18~T has little effect on the AFI transition
for nearly stoichiometric V$_2$O$_3$.

\begin{figure}[bt]
\centerline{
\psfig{file=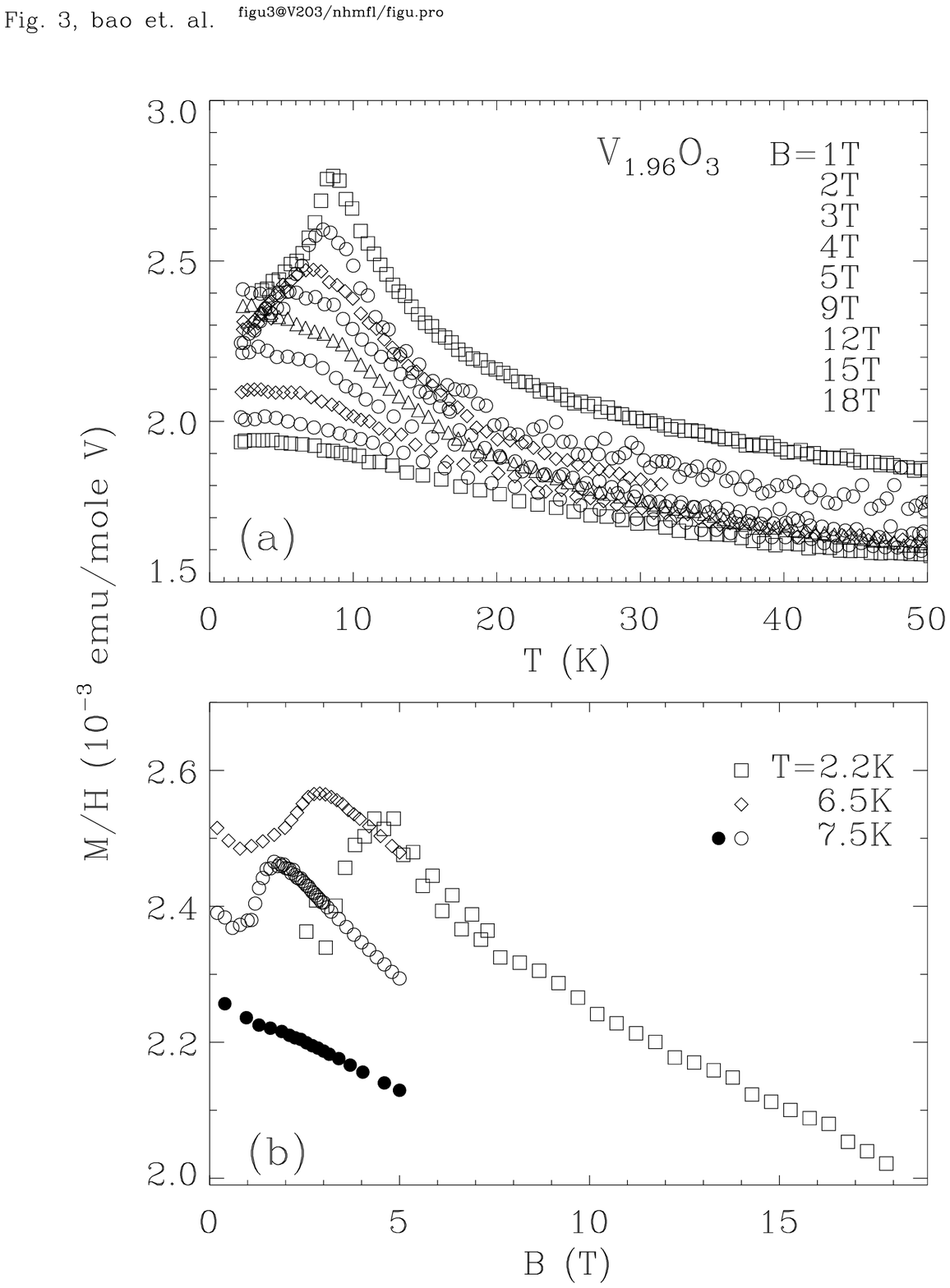,width=\columnwidth,angle=0,clip=}}
\caption{(a) Temperature dependence of M/H measured in fields from
1~T to 18~T (top to bottom). The SDW transition, indicated by the cusp, 
moves to lower temperature with increasing magnetic field. It is suppressed
for fields higher than 5~T.
(b) Magnetic field dependence of M/H at selected temperatures. Open
symbols represent results when a magnetic field is applied in the basal plane.
Filled symbol for field along the $c$ axis.}
\label{f_MvT}
\end{figure}

We now turn to magnetic field effect on the SDW transition.
Magnetization of V$_{1.96}$O$_3$ at various magnetic fields 
as a function of temperature is shown in Fig.~\ref{f_MvT}(a).
The cusp due to the second order SDW
phase transition near 8.7~K remains sharp at 1~T. 
With increasing magnetic field,
the cusp moves to lower temperature and becomes less well defined. 
Phase transition in this portion of the $T$-$B$ phase space
can be better determined with measurements of
magnetization as a function of the field at various fixed temperatures.
Some examples are shown in Fig.~\ref{f_MvT}(b).
These measurements complement well the constant field data in
Fig.~\ref{f_MvT}(a). 

\begin{figure}[bt]
\centerline{
\psfig{file=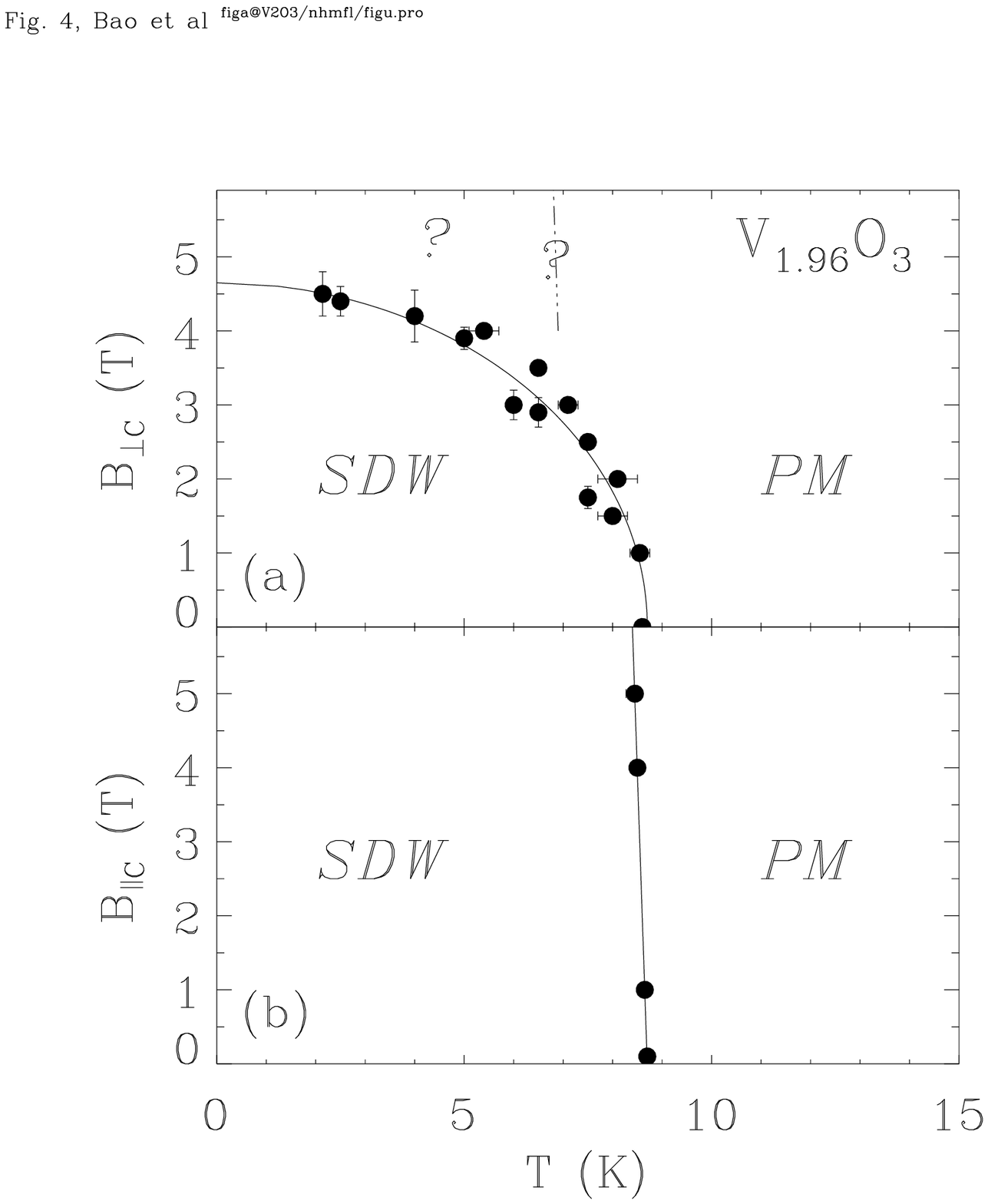,width=\columnwidth,angle=0,clip=}}
\caption{A T-B phase diagram for V$_{1.96}$O$_3$. 
The magnetic field direction is (a) perpendicular to and 
(b) parallel to the $c$ axis, respectively.}
\label{f_phs}
\end{figure}

Data points in Fig.~\ref{f_phs}(a) represent the peak positions
in the M vs T or M vs H curves, such as those in Fig.~\ref{f_MvT}.
They delineate the phase boundary for the transverse SDW phase.
The critical field to suppress the transverse SDW phase is 
extrapolated to be 4.6(3) T.

\begin{figure}[bt]
\centerline{
\psfig{file=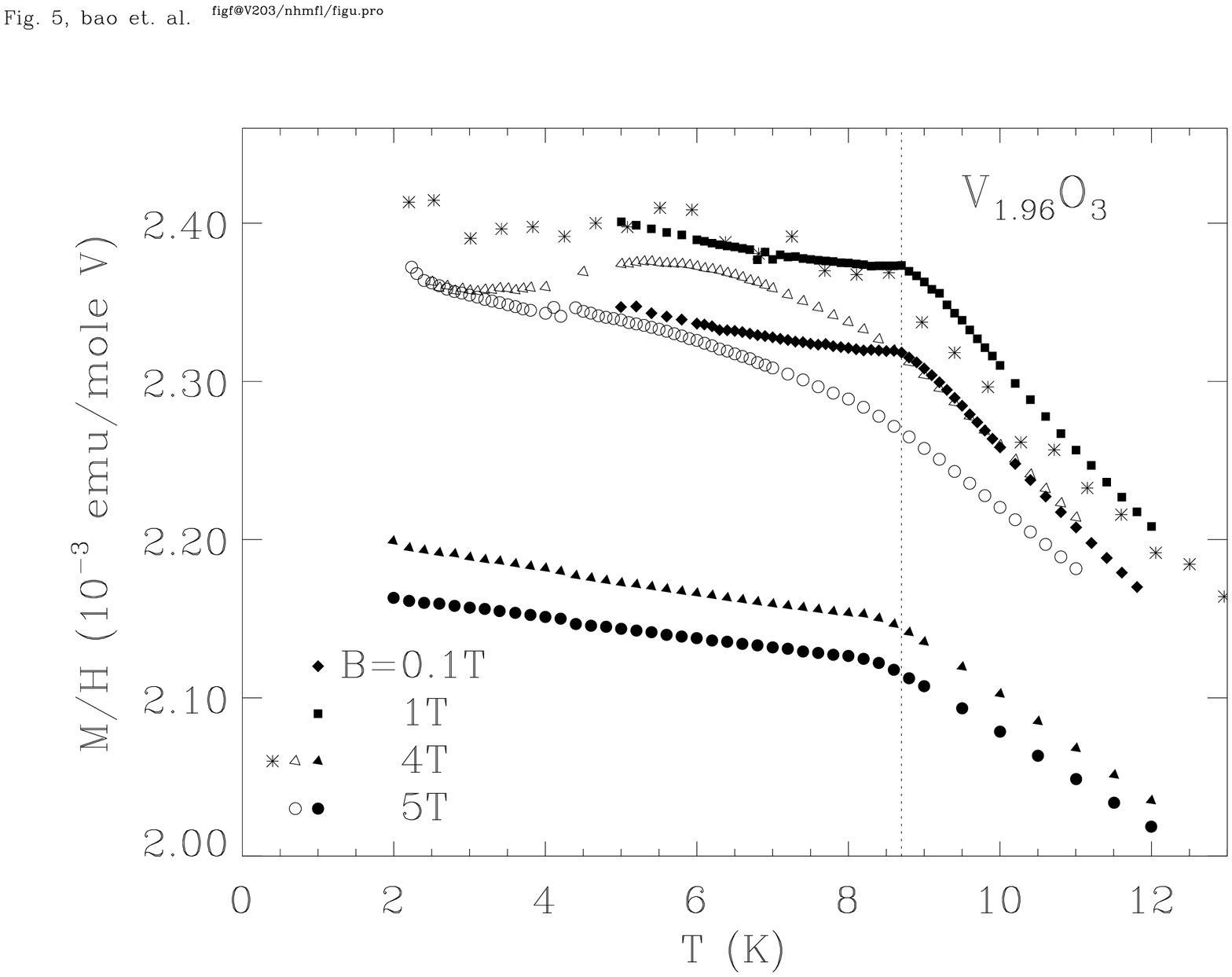,width=\columnwidth,angle=90,clip=}}
\caption{Temperature dependence of M/H measured with a SQUID magnetometer
with the field
applied along the $a^*$ axis (the open symbols) and the $c$ axis 
(the filled symbols). Data taken with a VSM at 4~T 
(star)
are also shown for comparison.
The vertical line indicates the zero-field N\'{e}el
temperature.
}
\label{f_dir}
\end{figure}

The transverse SDW state of V$_{1.96}$O$_3$ consists of spin spiral with
an incommensurate wave vector along the $c$ axis and 
spin direction rotating in the hexagonal basal plane\cite{bao93}. 
Thus, magnetic anisotropy is expected.
Additional measurements below 5~T, with sample orientation 
fixed with respect to the applied field,
were conducted using a SQUID magnetometer.
Magnetization measurements performed for the 
field applied along both the $c$ axis (filled symbols)
and $a^*$ axis (open symbols) are shown in Fig.~\ref{f_dir}. 
For a given magnetic field,
the $a^*$-axis magnetization is appreciably larger than the $c$-axis
magnetization, consistent with previous 
measurement\cite{bibyukkd}. 
Therefore, when the sample is allowed to rotate in
the magnetic field, as in the case of our VSM measurement with
the 20~T magnet, the measured response is predominantly the $a^*$-axis one.
This is confirmed by the similar magnetization data taken in 4~T using
VSM (star) and using SQUID with field applied along $a^*$ axis
(open triangle) in Fig.~\ref{f_dir}.
Thus, the phase diagram in Fig.~\ref{f_phs}(a) is
for the field in the rotating plane of the spin spiral, i.e.,
perpendicular to the $c$ axis.

The $c$-axis magnetization (filled symbols) in Fig.~\ref{f_dir}
shows a inflection around 8.7~K,
the zero field N\'{e}el point. This behavior is what is expected
of a antiferromagnet when the field direction is perpendicular to the spins.
While the magnitude of $M/H$
changes appreciably with magnetic field, the inflection point changes little.
This behavior in phase transition is in sharp contrast to that in
the $a$-axis magnetic response.
A different response is also clear in the M vs B curves at 7.5~K in
Fig.~\ref{f_MvT}(b). The cusp associating with the 
SDW phase transition appears
only when the field is perpendicular to the $c$ axis.
Phase boundary for $B$ parallel to the $c$ axis is shown in
Fig.~\ref{f_phs}(b).  There is little detectable effect on T$_N$.

The saturation moment for V$_2$O$_3$ in the AFI phase 
is 1.2$\mu_B$\cite{bibrmm}. The magnetic energy at 18~T,
1.2$\mu_B \times 18 {\rm T}=1.3$~meV, is only 9\% of ${\rm k_B T_C}$.
A much higher magnetic field may be needed to affect this first order
transition.

The saturation moments for V$_{1.96}$O$_3$ 
in the SDW phase is 0.15$\mu_B$\cite{bao93}. The magnetic energy for
V$_{1.96}$O$_3$ at the critical field, 
0.15$\mu_B \times 4.6 {\rm T}=0.04$~meV, is only 5\% of ${\rm k_B T_N}$.
In addition, the induced moment at 4.6 T and zero temperature
can be estimated using the value of
$M/H$ at low temperature from Fig.~\ref{f_MvT}(a):
$2.4\times 10^{-3}\times 46000=110$~emu/mole V $=0.02\mu_B$/V.
This is only a small fraction of the zero field staggered moment.
Therefore, it is unlikely that the SDW phase is replaced by a
field-induced ``ferromagnetic'' state above 4.6~T.
This is also consistent with the persistence of the SDW state at
5 T when the field is applied along the $c$ axis. In reference to
the T-B phase diagram of a common antiferromagnet, the phase
above 4.6 T in Fig.~\ref{f_phs}(a) could be a spin-flop state.
In this case, the B$\cdot$M is comparable to the anisotropic energy,
which can be a small fraction of ${\rm k_B T_N}$.
It remains to be seen, using, e.g., neutron diffraction, what happens to
the incommensurate magnetic structure of the transverse SDW at
high magnetic field and low temperature. 
A field-induced incommensurate-commensurate
transition has been observed in the Dzyaloshinskii-Moriya 
type\cite{DzyMo} incommensurate magnets, such as 
the itinerant MnSi\cite{mnsi_phs} and 
localized Ba$_2$CuGe$_2$O$_7$\cite{andrey}, while the 
incommensurability is expected to be locked by 
Fermi surface in Lomer-Overhauser type\cite{sdw_lom}
incommensurate magnets.

Recently, magnetotransport measurements have been conducted by 
Klimm et al.\cite{klimm} up to 12~T. Anomalies in their data may be related
to the phase transition uncovered here. However, the orientation dependence in 
their data has yet to be
understood in term of the orientation dependence we find in this work.

\acknowledgments

We thank J. L. Sarrao for help in Laue x-ray diffraction.
Work at Los Alamos was performed under the auspices
of the U.S. Department of Energy. Work performed at the
NHMFL was sponsored by the NSF,
with additional support from the State of Florida and the U.S. Department of 
Energy.


\end{document}